\documentclass[aps,prl,onecolumn,superscriptaddress]{revtex4}

\usepackage{graphicx}
\usepackage{amsmath}
\usepackage{amsfonts}
\usepackage{bm}% bold math
\usepackage{color}
\usepackage{rotating}
\usepackage{array}

\begin{document}

\title{Quantum erasure with causally disconnected choice}

\author{Xiao-song Ma}
\email{Xiaosong.Ma@univie.ac.at}
\affiliation{Institute for Quantum Optics and Quantum Information (IQOQI), Austrian
Academy of Sciences, Boltzmanngasse 3, A-1090 Vienna, Austria}
\affiliation{Vienna Center for Quantum Science and Technology, Faculty of Physics, University of Vienna, Boltzmanngasse 5, A-1090 Vienna,
Austria}
\affiliation{Present address: Department of Electrical Engineering, Yale University, New Haven, CT 06511, USA}
\author{Johannes Kofler}
\affiliation{Institute for Quantum Optics and Quantum Information (IQOQI), Austrian
Academy of Sciences, Boltzmanngasse 3, A-1090 Vienna, Austria}
\affiliation{Present address: Max Planck Institute of Quantum Optics, Hans-Kopfermann-Str.\ 1, 85748 Garching/Munich, Germany}
\author{Angie Qarry}
\affiliation{Institute for Quantum Optics and Quantum Information (IQOQI), Austrian
Academy of Sciences, Boltzmanngasse 3, A-1090 Vienna, Austria}
\affiliation{Vienna Center for Quantum Science and Technology, Faculty of Physics, University of Vienna, Boltzmanngasse 5, A-1090 Vienna,
Austria}
\author{Nuray Tetik}
\affiliation{Institute for Quantum Optics and Quantum Information (IQOQI), Austrian
Academy of Sciences, Boltzmanngasse 3, A-1090 Vienna, Austria}
\affiliation{Vienna Center for Quantum Science and Technology, Faculty of Physics, University of Vienna, Boltzmanngasse 5, A-1090 Vienna,
Austria}
\author{Thomas Scheidl}
\affiliation{Institute for Quantum Optics and Quantum Information (IQOQI), Austrian
Academy of Sciences, Boltzmanngasse 3, A-1090 Vienna, Austria}
\author{Rupert Ursin}
\affiliation{Institute for Quantum Optics and Quantum Information (IQOQI), Austrian
Academy of Sciences, Boltzmanngasse 3, A-1090 Vienna, Austria}
\author{Sven Ramelow}
\affiliation{Institute for Quantum Optics and Quantum Information (IQOQI), Austrian
Academy of Sciences, Boltzmanngasse 3, A-1090 Vienna, Austria}
\affiliation{Vienna Center for Quantum Science and Technology, Faculty of Physics, University of Vienna, Boltzmanngasse 5, A-1090 Vienna,
Austria}
\author{Thomas Herbst}
\affiliation{Faculty of Physics, University of Vienna, Boltzmanngasse 5, A-1090 Vienna,
Austria}
\affiliation{Institute for Quantum Optics and Quantum Information (IQOQI), Austrian
Academy of Sciences, Boltzmanngasse 3, A-1090 Vienna, Austria}
\author{Lothar Ratschbacher}
\affiliation{Institute for Quantum Optics and Quantum Information (IQOQI), Austrian
Academy of Sciences, Boltzmanngasse 3, A-1090 Vienna, Austria}
\affiliation{Vienna Center for Quantum Science and Technology, Faculty of Physics, University of Vienna, Boltzmanngasse 5, A-1090 Vienna,
Austria}
\affiliation{Present address: Cavendish Laboratory, University of Cambridge, JJ Thomson Avenue, Cambridge CB3 0HE, United Kingdom}
\author{Alessandro Fedrizzi}
\affiliation{Institute for Quantum Optics and Quantum Information (IQOQI), Austrian
Academy of Sciences, Boltzmanngasse 3, A-1090 Vienna, Austria}
\affiliation{Vienna Center for Quantum Science and Technology, Faculty of Physics, University of Vienna, Boltzmanngasse 5, A-1090 Vienna,
Austria}
\affiliation{Present address: Centre for Engineered Quantum Systems and Centre for Quantum Computer and Communication Technology, School of Mathematics and Physics, University of Queensland, Brisbane 4072, Australia}
\author{Thomas Jennewein}
\affiliation{Institute for Quantum Optics and Quantum Information (IQOQI), Austrian
Academy of Sciences, Boltzmanngasse 3, A-1090 Vienna, Austria}
\affiliation{Present address: Institute for Quantum Computing and Department of Physics and Astronomy, University of Waterloo, 200 University Avenue West, Waterloo, ON, N2L 3G1, Canada}
\author{Anton Zeilinger}
\email{Anton.Zeilinger@univie.ac.at}
\affiliation{Institute for Quantum Optics and Quantum Information (IQOQI), Austrian
Academy of Sciences, Boltzmanngasse 3, A-1090 Vienna, Austria}
\affiliation{Vienna Center for Quantum Science and Technology, Faculty of Physics, University of Vienna, Boltzmanngasse 5, A-1090 Vienna, Austria}
\affiliation{Faculty of Physics, University of Vienna, Boltzmanngasse 5, A-1090 Vienna, Austria}

\begin{abstract}
The counterintuitive features of quantum physics challenge many common-sense assumptions. In an interferometric quantum eraser experiment, one can actively choose whether or not to erase which-path information, a particle feature, of one quantum system and thus observe its wave feature via interference or not by performing a suitable measurement on a distant quantum system entangled with it. In all experiments performed to date, this choice took place either in the past or, in some delayed-choice arrangements, in the future of the interference. Thus in principle, physical communications between choice and interference were not excluded. Here we report a quantum eraser experiment, in which by enforcing Einstein locality no such communication is possible. This is achieved by independent active choices, which are space-like separated from the interference. Our setup employs hybrid path-polarization entangled photon pairs which are distributed over an optical fiber link of 55 m in one experiment, or over a free-space link of 144 km in another. No naive realistic picture is compatible with our results because whether a quantum could be seen as showing particle- or wave-like behavior would depend on a causally disconnected choice. It is therefore suggestive to abandon such pictures altogether.
\end{abstract}

\maketitle

Wave-particle duality is a well-known manifestation of the more general complementarity principle in quantum physics~\cite{Bohr1949}. Several single-photon experiments~\cite{Clauser1974, Grangier1986, Dopfer1998, Zeilinger2000, Zeilinger2005} confirmed both the wave and the particle nature of light. Another manifestation of complementarity is that the position and linear momentum of individual particles cannot be well-defined together as highlighted in Heisenberg's uncertainty relation~\cite{Heisenberg1927}. Based on the concept of the Heisenberg microscope~\cite{Heisenberg1927}, von Weizs\"{a}cker~\cite{Weizsacker1931,Weizsacker1941} discussed the gedanken experiment in which a photon interacts with an electron.  In today's language, after the interaction the photon and the electron are in an entangled state~\cite{Einstein1935,Schroedinger1935}, in which their positions and momenta are strongly correlated. Therefore, different complementary measurements on the photon allow choosing whether the electron acquires a well-defined position or a well-defined momentum, respectively. According to Bohr, ``it obviously can make no difference as regards observable effects [...] whether our plans of constructing or handling the instruments are fixed beforehand or whether we prefer to postpone the completion of our planning until a later moment [...]''~\cite{Bohr1949}.

Wheeler later proposed an experiment on wave-particle duality in which the paths of a single photon, coming from a distant star, form a very large interferometer~\cite{Wheeler1978, Wheeler1984}. Inserting or not inserting a beam splitter at the end of the interferometer's paths will allow to either observe interference (wave behavior) or acquire path information (particle behavior), respectively. Wheeler proposed to delay the choice whether or not to insert the beam splitter until the very last moment of the photon's travel inside the interferometer. This rules out the possibility that the photon knew the configuration beforehand and adapted its behavior accordingly~\cite{Mittelstaedt1987a,Greenstein2005}. He then pointed out the seemingly paradoxical situation that it depends on the experimenter's delayed choice whether the photon behaved as a particle or a wave. In Wheeler's words: ``We, now, by moving the mirror in or out have an unavoidable effect on what we have a right to say about the \emph{already} past history of that photon"~\cite{Wheeler1984}.  Since then, Wheeler's proposal has led to several experimental studies with single-photon interference~\cite{Hellmuth1985, Alley1987, Hellmuth1987, Baldzuhn1989, Lawson-Daku1996, Jacques2007, Jacques2008}, which provided increasingly sophisticated demonstrations of the wave-particle duality of single quanta, even in a delayed-choice configuration.

Scully and Dr\"{u}hl proposed the so-called quantum eraser~\cite{Scully1982, Scully1991}, in which maximally entangled atom-photon states were studied. In~\cite{Scully1991}, the atoms, which can be interpreted as the `system', are sent through a double slit. Each atom spontaneously emits a photon, which can be regarded as the `environment', carrying welcher-weg (which-path) information on which of the two slits the atom takes. No interference pattern of atoms will be obtainable after the double slit, if one ignores the presence of the photons, since every photon carries the welcher-weg information about the corresponding atom. The presence of path information anywhere in the universe is sufficient to prohibit any possibility of interference. It is irrelevant whether a future observer might decide to acquire it. The mere possibility is enough. In other words, the atoms' path states alone are not in a coherent superposition due to the atom-photon entanglement.

If the observer measures the photons, his choice of the type of measurement decides whether the atoms can be described by a wave or a particle picture. Firstly, when the photons are measured in a way that reveals welcher-weg information of the atoms, the atoms do not show interference, not even conditionally on the photons' specific measurement results. Secondly, if the photons are measured such that this irrevocably erases any welcher-weg information about the atoms, then the atoms will show perfect but distinct interference patterns, which are each other's complement and are conditioned on the specific outcomes of the photons' measurements. These two scenarios illustrate a further manifestation of the complementarity principle, in addition to the wave-particle duality. There is a trade-off between acquiring the atoms' path information or their interference pattern via complementary measurements on the photons and not on the atoms themselves. A continuous transition between these two extreme situations exists, where partial welcher-weg information and interference patterns with reduced visibility can be obtained~\cite{Wootters1979,Mittelstaedt1987b}.

The authors of refs.~\cite{Scully1982, Scully1991} proposed to combine the delayed-choice paradigm with the quantum eraser concept. Since the welcher-weg information of the atoms is carried by the photons, the choice of measurement of the photons---either revealing or erasing the atoms' welcher-weg information---can be delayed until ``long after the atoms have passed" the photon detectors at the double slit~\cite{Scully1991}. The later measurement of the photons `decides' whether the atoms can show interference or not even after the atoms have been detected. This seemingly counter-intuitive situation comes from the fact that in a bipartite quantum state the observed correlations are independent of the space-time arrangement of the measurements on the individual systems.  Thereby, their proposed scheme significantly extended the concept of the single-photon delayed-choice gedanken experiment as introduced by Wheeler and stimulated a lot of theoretical and experimental research~\cite{Englert2000, Kwiat2004, Aharonov2005, Eichmann1993, Kim2000, Walborn2002}. Also, the proposal~\cite{Peres2000} and the experimental realizations of delayed-choice entanglement swapping~\cite{Jennewein2001,Sciarrino2002,Jennewein2005,Ma2012} were reported. Recently, a quantum delayed-choice experiment was proposed~\cite{Ionicioiu2011} and realized{~\cite{Tang2012,Peruzzo2012}}. During the course of writing the present manuscript, the experiment in~\cite{Kaiser2012} reported space-like separation between the outcomes of all measurements. In addition we employ ultra-fast switching as well as precisely timed random setting choices, to conclusively ensure the space-like separation of all relevant events (setting choices, setting implementations, measurements). This also also made possible many different space-time scenarios.

Here, for the first time, we propose and experimentally demonstrate a quantum eraser under enforced Einstein locality. The locality condition imposes that if ``two systems no longer interact, no real change can take place in the second system in consequence of anything that may be done to the first system"~\cite{Einstein1935}. Operationally, to experimentally realize a quantum eraser under Einstein locality conditions, the erasure event of welcher-weg information has to be relativistically space-like separated from the whole passage of the interfering system through the interferometer including its final registration. This means that in any and all reference frames no subluminal or luminal physical signal can travel from one event to the other and causally influence it. Implementing Einstein locality thus implies a significant step in the history of quantum eraser experiments.

\begin{figure}
    \includegraphics[width=0.5\textwidth]{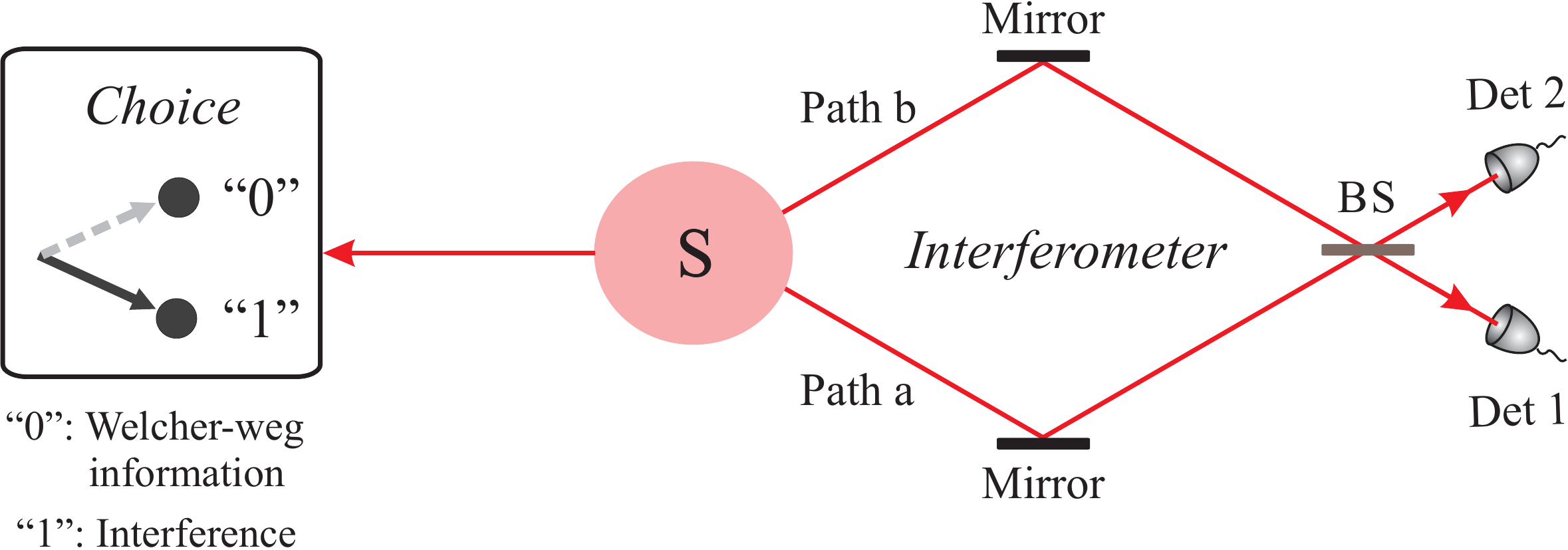}
    \caption{The concept of our quantum eraser under Einstein locality conditions. The hybrid entangled photon-pair source, labeled as S, emits path-polarization entangled photon pairs. The system photons are propagating through an interferometer on the right side and the environment photons are subject to polarization measurements on the left side. The choices to acquire welcher-weg information or to obtain interference of the system photons are made under Einstein locality, so that there are no causal influences between the system photons and the environment photons.}\label{concept}
\end{figure}

The concept of our quantum eraser is illustrated in Fig.\ \ref{concept}. We produce hybrid entangled photon pairs~\cite{Ma2009}, with entanglement between two different degrees of freedom, namely the path of one photon, denoted as the system photon, and the polarization of the other photon, denoted as the environment photon. The system photon is sent to an interferometer, and the environment photon is sent to a polarization analyzer which performs a measurement according to a causally disconnected choice (with respect to the interferometer-related events). Analogous to the original proposal of the quantum eraser~\cite{Scully1982, Scully1991}, the environment photon's polarization carries welcher-weg information of the system photon due to the entanglement between the two photons.  Depending on which polarization basis the environment photon is measured in, we are able to either acquire welcher-weg information of the system photon and observe no interference, or erase welcher-weg information and observe interference. In the latter case, it depends on the specific outcome of the environment photon which one out of two different interference patterns the system photon is showing. Basic results of our work have been reported in~\cite{Ma2007,Ma2008}, and more information can be found in~\cite{Ma2010}.

\begin{figure}
    \includegraphics[width=0.8\textwidth]{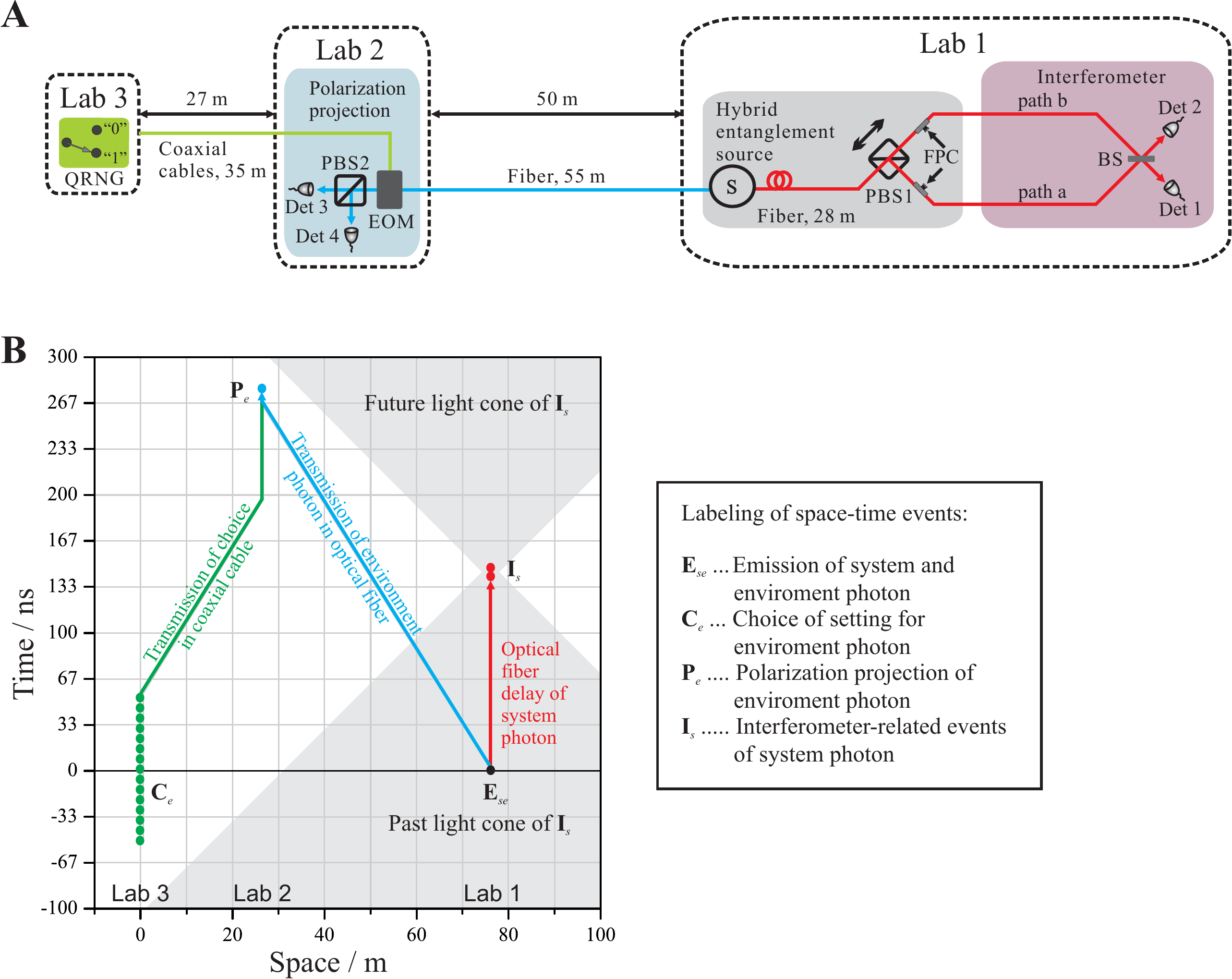}
    \caption{\textbf{A}. Scheme of the Vienna experiment: In Lab1, the source (S) emits polarization entangled photon pairs, each consisting of a system and an environment photon, via type-II spontaneous parametric down-conversion. Good spectral and spatial mode overlap is achieved by using interference filters with 1 nm bandwidth and by collecting the photons into single-mode fibres.  The polarization entangled state is subsequently converted into a hybrid entangled state with a polarizing beam splitter (PBS1) and two fibre polarization controllers (FPC).  The interferometric measurement of the system photon is performed with a single-mode fibre beam splitter (BS) with a path length of 2 m, where the relative phase between path a and path b is adjusted by moving PBS1's position with a piezo-nanopositioner. The polarization projection setup of the environment photon consists of an electro-optical modulator (EOM) and another polarizing beam splitter (PBS2).  Both photons are detected by silicon avalanche photodiodes (Det~1-4).  The choice is made with a Quantum Random Number Generator (QRNG)~\cite{Jennewein2000}. \textbf{B}. Space-time diagram.  The choice-related events \textbf{C}$_{e}$ and the polarization projection of the environment photon \textbf{P}$_{e}$ are space-like separated from all events of the interferometric measurement of the system photon \textbf{I}$_{s}$.  Additionally, the events \textbf{C}$_{e}$ are also space-like separated from the emission of the entangled photon pair from the source \textbf{E}$_{se}$.  The shaded areas are the past and the future light cones of events \textbf{I}$_{s}$.  This ensures that Einstein locality is fulfilled. For details, see main text and supplementary information.}\label{setup}
\end{figure}

\begin{figure}[t!]
    \includegraphics[width=0.70\textwidth]{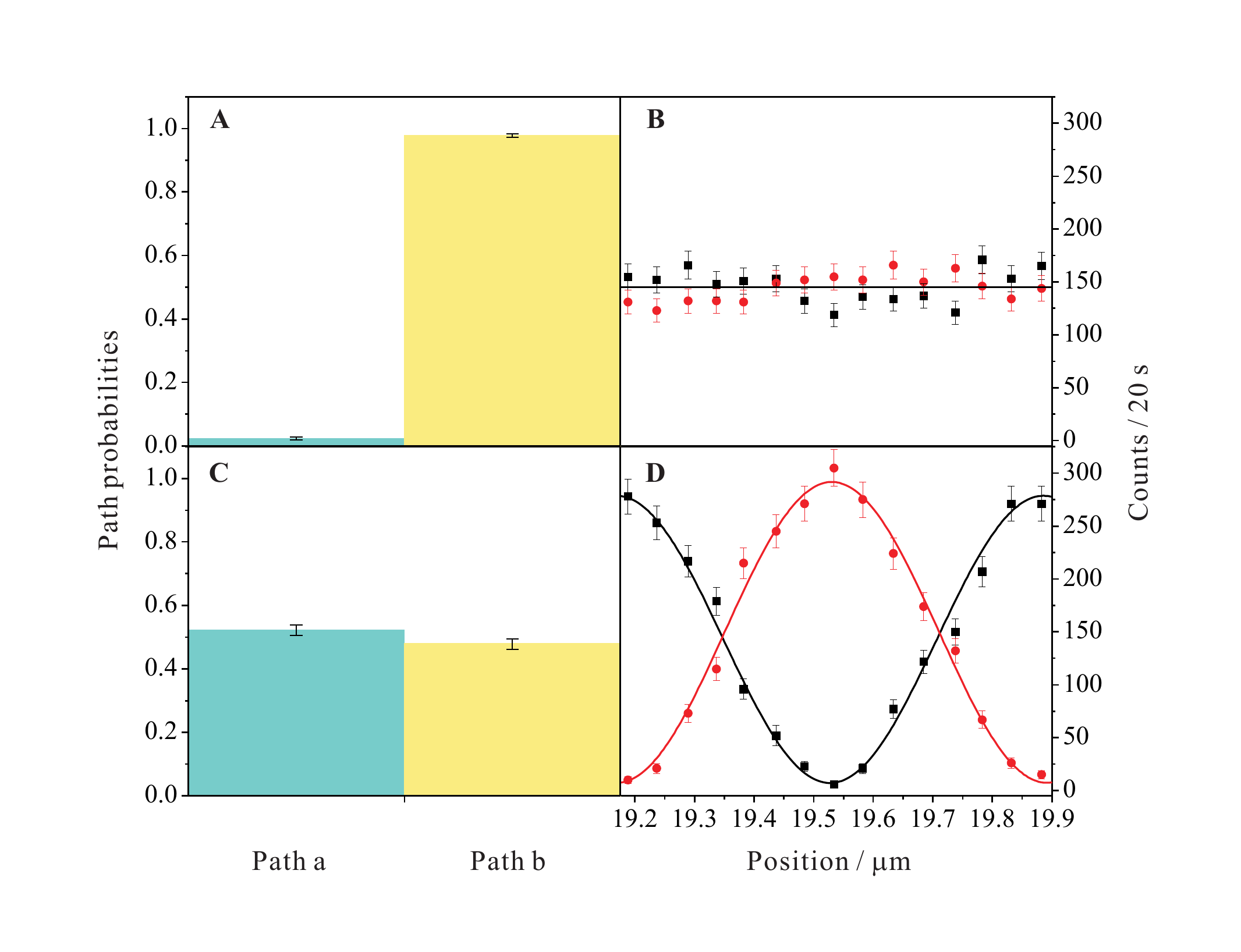}\vspace{-0.75cm}
    \caption{Experimental results. Top: When measurement (i) is performed (EOM is off), the detection of the environment photon in the state ${\left| \textrm{V} \right\rangle} _{e} $ reveals the `welcher-weg' information of the system photon, being confirmed by measuring the counts of Det~1 and Det~2 conditional on the detection of the environment photon in Det~4. \textbf{A}. We obtain that the system photon propagates through path a and path b with probabilities 0.023(5) (cyan) and 0.978(5) (yellow), respectively. The integration time is about 120 s. As a consequence of revealing welcher-weg information, phase insensitive counts are obtained.  The mean value of the counts is indicated with a black line, as shown in \textbf{B}. Bottom: When measurement (ii) is performed (EOM on), detection of the environment photon in ${\left| \textrm{R} \right\rangle} _{e} $ erases the welcher-weg information of the system photon. \textbf{C}. The probabilities of the system photon propagating through path a and path b are 0.521(16) (cyan) and 0.478(16) (yellow), respectively. The integration time is about 120 s. Because welcher-weg information is irrevocably erased, two oppositely modulated sinusoidal interference fringes with average visibility 0.951(18) show up as a function of the position change of PBS1, as shown in \textbf{D}. The error bars stand for $\pm$1 standard deviation and are given by Poissonian statistics.}\label{results}
\end{figure}

To test the quantum eraser concept under various spatio-temporal situations, we performed several experiments demonstrating the quantum eraser under Einstein locality on two different length scales.  In the first experiment performed in Vienna in 2007, the environment photon is sent away from the system photon via a 55~m long optical fibre. In the second experiment performed on the Canary Islands in 2008, they are separated by 144 km and connected via a free-space link. The scheme of our Vienna experiment is shown in Fig.\ \ref{setup}\textbf{A}.  First, we prepare a polarization-entangled state~\cite{Kwiat1995}: ${({\left| \textrm{H} \right\rangle} _{s} {\left| \textrm{V} \right\rangle} _{e} +{\left| \textrm{V} \right\rangle} _{s} {\left| \textrm{H} \right\rangle} _{e} )\mathord{\left/ {\vphantom {({\left| \textrm{H} \right\rangle} _{s} {\left| \textrm{V} \right\rangle} _{e} +{\left| \textrm{V} \right\rangle} _{s} {\left| \textrm{H} \right\rangle} _{e} ) \sqrt{2} }} \right. \kern-\nulldelimiterspace} \sqrt{2} } $, where ${\left| \textrm{H} \right\rangle}$ and ${\left| \textrm{V} \right\rangle} $ denote quantum states of horizontal and vertical linear polarization, and \textit{s} and \textit{e} index the system and environment photon, respectively.  The orthogonal polarization states of the system photon are coherently converted into two different interferometer path states ${\left| \textrm{a} \right\rangle} _{s} $ and ${\left| \textrm{b} \right\rangle} _{s} $ via a polarizing beam splitter and two fiber polarization controllers. This approximately generates the hybrid entangled state (for details on imperfections and reduced state purity see supplementary information)~\cite{Ma2009}:
\begin{equation} \label{GrindEQ__1_}
{\left| \Psi _{\textrm{hybrid}}  \right\rangle} _{se} ={\tfrac{1}{\sqrt{2} }} ({\left| \textrm{b} \right\rangle} _{s} {\left| \textrm{V} \right\rangle} _{e} +{\left| \textrm{a} \right\rangle} _{s} {\left| \textrm{H} \right\rangle} _{e} ).
\end{equation}
The environment photon thus carries welcher-weg information about the system photon. Therefore, we are able to perform two complementary polarization projection measurements on the environment photon and acquire or erase welcher-weg information of the system photon, respectively. (i) We project the environment photon into the H/V basis, which reveals welcher-weg information of the system photon and no interference can be observed; (ii) We project the environment photon into the R/L basis (with ${\left| \textrm{R} \right\rangle} =({\left| \textrm{H} \right\rangle} +i{\left| \textrm{V} \right\rangle} )/\sqrt{2} $ and ${\left| \textrm{L} \right\rangle} =({\left| \textrm{H} \right\rangle} -i{\left| \textrm{V} \right\rangle} )/\sqrt{2} $) of left and right circular polarization states, which erases welcher-weg information.  Contrary to the first case, the detection of the environment photon in polarization R (or L) results in a coherent superposition with equal probabilities for the states ${\left| \textrm{a} \right\rangle} _{s} $ and ${\left| \textrm{b} \right\rangle} _{s} $, as Eq.~\eqref{GrindEQ__1_} can be rewritten as
\begin{equation} \label{GrindEQ__3_}
{\left| \Psi _{\textrm{hybrid}}  \right\rangle} _{se}^{} ={\tfrac{1}{2}} [({\left| \textrm{a} \right\rangle} _{s} +i{\left| \textrm{b} \right\rangle} _{s} ){\left| \textrm{L} \right\rangle} _{e} +({\left| \textrm{a} \right\rangle} _{s} -i{\left| \textrm{b} \right\rangle} _{s} ){\left| \textrm{R} \right\rangle} _{e} ].
\end{equation}
In case (ii), the polarization of the environment photon (either R or L) carries information about the relative phase between paths a and b of the system photon. This gives rise to complementary interference patterns (fringes or antifringes). The cases (i) and (ii) show that the which-path information and the fringe-antifringe information are equally fundamental. Note that similar setups have been proposed in refs.\ \cite{Grangier1986a, Ballentine1998, Kwiat2004}.
\begin{figure}[ht!]
    \includegraphics[width=0.42\textwidth]{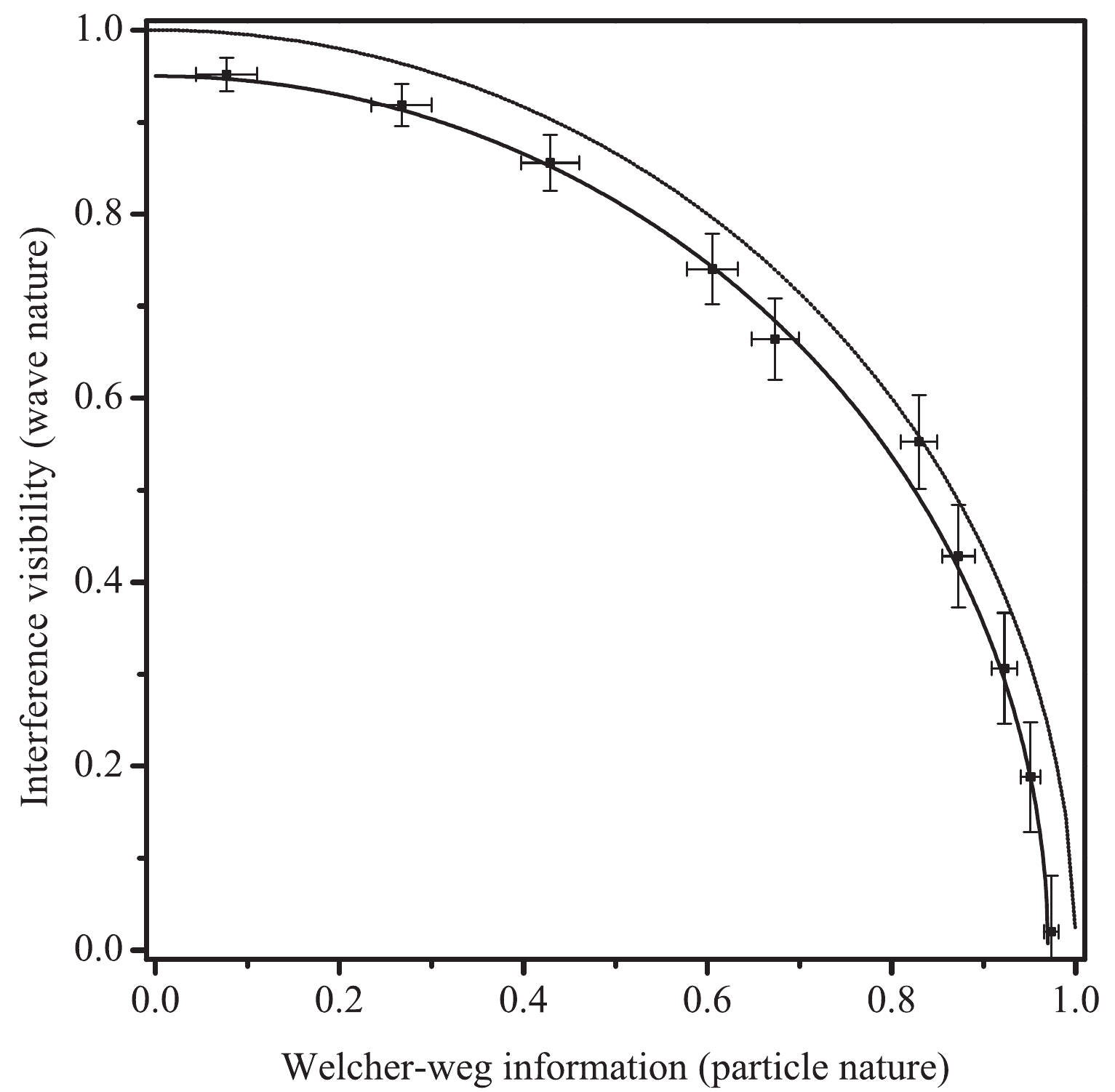}
    \caption{Experimental test of the complementarity inequality under Einstein locality, manifested by a trade-off of the welcher-weg information parameter and the interference visibility.  We vary the polarization projection basis of the environment photon via adjusting the applied voltage of the EOM. Note that the leftmost and the rightmost data points correspond to Fig.\ \ref{results} \textbf{A}-\textbf{B} and \textbf{C}-\textbf{D}, respectively. The dotted line is the ideal curve from the saturation of Ineq.\ \eqref{GrindEQ__2_}. The solid line, $\mathcal{V}=0.95\sqrt{1-\left(\mathcal{I}/0.97\right)^{2}}$, is the estimation from the actual experimental imperfections, which are measured independently. The error bars stand for $\pm$1 standard deviation and are given by Poissonian statistics.}\label{com}
\end{figure}

The following events are important and should be identified before the discussion of the space-time diagram: \textbf{E}$_{se}$ is the emission of both the system photon and the environment photon from the source, \textbf{C}$_{e}$ is the choice of the polarization measurement basis of the environment photon, \textbf{P}$_{e}$ is the polarization projection of the environment photon, and \textbf{I}$_{s}$ are all events related to the system photon inside the interferometer including its entry into, its propagation through, and its exit from the interferometer.

In order to guarantee Einstein locality for a conclusive test, any causal influence between choice \textbf{C}$_{e}$ and projection \textbf{P}$_{e}$ of the environment photon on one hand and interferometer-related events \textbf{I}$_{s}$ of the system photon on the other has to be ruled out. Operationally, we require space-like separation of \textbf{C}$_{e}$, \textbf{P}$_{e}$ with respect to \textbf{I}$_{s}$ (see Fig.\ \ref{setup}\textbf{B}). All this is achieved by setting up the respective experimental apparatus in three distant labs. The choice is performed by a quantum random number generator (QRNG). See supplementary information for details. Its working principle is based on the intrinsically random detection events of photons behind a balanced beam splitter~\cite{Jennewein2000}.

Note that our setup also excludes any dependence between the choice and the photon pair emission (``freedom of choice"~\cite{Bell2004,Scheidl2010}), because we locate the source and QRNG in two separate labs such that space-like separation between the events \textbf{C}$_{e}$ and \textbf{E}$_{se}$ is ensured. In~\cite{Kim2000}, the choice is made passively by the environment photon itself and therefore is situated in the future light cone of both the emission of the photon pair and the measurement event of the system photon.  Therefore, it is in principle conceivable that the emission event and system photon measurement event can influence the choice, which then only appears to be free or random.

In Fig.\ \ref{results}, we present the experimental results for measurements of the system photon conditioned on the detection of the environment photon with Det~4.  In Fig.\ \ref{results}\textbf{A}, the probabilities that the system photon takes path a or b are shown when measurement (i), i.e.\ projection of the environment photon into the H/V basis and thus acquiring welcher-weg information, is performed.  When the environment photon is subjected to measurement (i) and detected to have polarization V, the probability that the system photon propagates through path a is $P\left(\textrm{a}\left|\textrm{V} \right. \right)$ = 0.023(5), which is determined by blocking path b and summing up the coincidence counts over 120 s between both interferometer detectors and V detectors. Likewise, we find that the probability for propagation through path b is $P\left(\textrm{b}\left|\textrm{V} \right. \right)$ = 0.978(5). In order to quantify the amount of welcher-weg information acquired, we use the so-called welcher-weg information parameter~\cite{Wootters1979, Greenberger1988, Englert1996, Englert2000}, $\mathcal{I}_{\textrm{(i)}}=|P(\textrm{a}|\textrm{V})-P(\textrm{b}|\textrm{V})|$. The value 0.955(7) of the parameter $\mathcal{I}_{\textrm{(i)}}$ reveals almost full welcher-weg information of the system photon.  As a consequence, when the relative phase between path a and b is scanned, no interference pattern is observed as shown in Fig.\ \ref{results}\textbf{B}. We integrate 20~s for each data point.

\begin{figure}[t!]
    \begin{center}
    \includegraphics[width=0.65\textwidth]{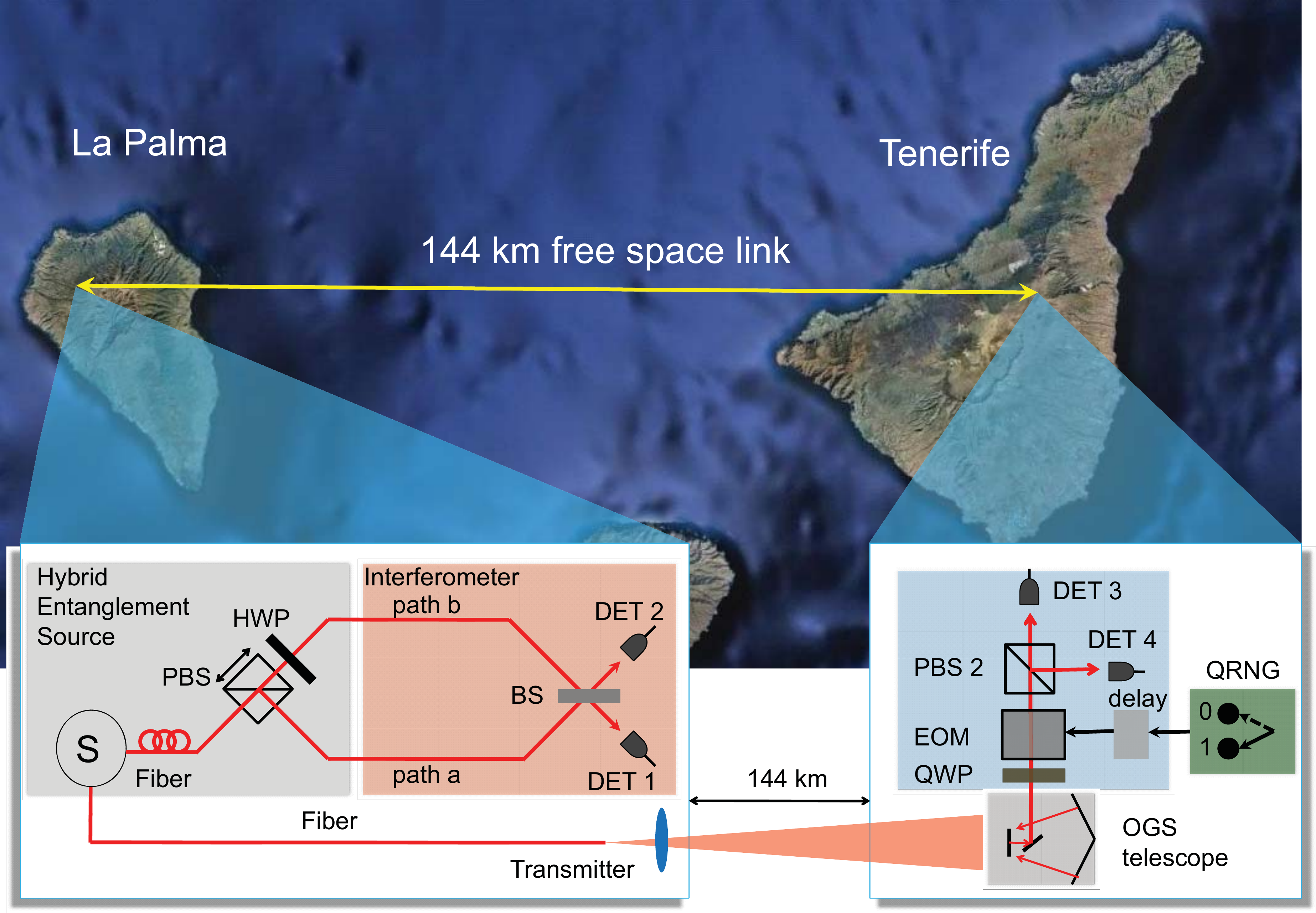}\\
    \caption{Satellite image of the Canary Islands of Tenerife and La Palma and overview of the experimental setup (Google Earth). The two labs are spatially separated by about 144 km.  In La Palma, the source (S) emits polarization entangled photon pairs, which subsequently are converted to a hybrid entangled state with a polarizing beam splitter (PBS1) and a half-wave plate oriented at 45$^\circ$.  The interferometric measurement of the system photon is done with a free-space beam splitter (BS), where the relative phase between path a and path b is adjusted by moving PBS1's position with a piezo-nanopositioner. The total path length of this interferometer is about 0.5 m. The projection setup consists of a quarter-wave plate (QWP), an electro optical modulator (EOM) and a polarizing beam splitter (PBS2), which together project the environment photon into either the H/V or +/-- basis (with ${\left| \textrm{+} \right\rangle} =({\left| \textrm{H} \right\rangle} +{\left| \textrm{V} \right\rangle} )/\sqrt{2} $ and ${\left| \textrm{--} \right\rangle} =({\left| \textrm{H} \right\rangle} -{\left| \textrm{V} \right\rangle} )/\sqrt{2} $). Both the system photon and the environment photon are detected by silicon avalanche photodiodes (DET~1-4).  A Quantum Random Number Generator (QRNG) defines the choice for the experimental configuration fast and randomly.  A delay card is used to adjust the relative time between the choice event and the other events.  Independent data registration is performed by individual time-tagging units on both the system and environment photon sides.  The time bases on both sides are established by the Global Positioning System (GPS).}\label{canarysetup}
    \end{center}
\end{figure}
On the other hand, when the environment photon is subjected to measurement (ii), i.e.\ projection of the environment photon into L/R basis, the welcher-weg information is irrevocably erased. When it is detected to have polarization R, we obtain the probabilities of the system photon propagating through path a, $P(\textrm{a}|\textrm{R})$ = 0.521(16), and through path b, $P(\textrm{b}|\textrm{R})$ = 0.478(16) (Fig.\ \ref{results}\textbf{C}).  In this case, $\mathcal{I}_{\textrm{(ii)}}$, defined as $\mathcal{I}_{\textrm{(ii)}}=|P(\textrm{a}|\textrm{R})-P(\textrm{b}|\textrm{R})|$, has the small value 0.077(22).  Accordingly, interference shows up with the visibility of $\mathcal{V}_{\textrm{(ii)}} = 0.951(18)$ as shown in Fig.\ \ref{results}\textbf{D}, where we integrate 20~s for each data point. This visibility is defined as $\mathcal{V}=(C_{\max } -C_{\min })/(C_{\max }+C_{\min })$, where $C_{\max } $ and $C_{\min } $ are the maximum and minimum counts of the system photon conditioned on the detection of the environment photon with Det~4. If the environment photon is detected to have polarization L, a $\pi$-phase shifted interference pattern of the system photons shows up. These results together with the space-time arrangement of our experiment conclusively confirm the acausal nature of the quantum eraser concept.

Quantum mechanics predicts the correlations of the measurement results to be invariant upon change of the specific space-time arrangement. We therefore realized another five qualitatively different space-time scenarios which are summarized in the Supplementary Information. All results obtained indeed agree with the expectation within statistical errors.

In order to quantitatively demonstrate the quantum eraser and the complementarity principle under Einstein locality, we employ a bipartite complementarity inequality~\cite{Wootters1979, Greenberger1988, Englert1996}, namely,
\begin{equation} \label{GrindEQ__2_}
\mathcal{I}^{2} +\mathcal{V}^{2} \le 1,
\end{equation}
which is an extension of the single-particle complementarity inequality (experimentally verified in ref.\ \cite{Jacques2008}). Here $\mathcal{I}$ and $\mathcal{V}$ are the parameters for two particles, as defined above. In an ideal experimental arrangement, inequality~\eqref{GrindEQ__2_} is saturated. Under Einstein locality, we measure $\mathcal{I}$ and $\mathcal{V}$ in sequential experimental runs as a function of the applied voltage of the EOM, which changes the polarization projection basis of the environment photon. Hence, we obtain a continuous transition between measurements (i) and (ii) and thus between particle and wave features. For each measurement, according to the QRNG output, the voltage of the EOM is randomly and rapidly switched between 0 and a definite value.  The results are shown in Fig.\ \ref{com}.  The dashed line is the ideal curve, where $\mathcal{I}^{2} +\mathcal{V}^{2} = 1$.  The solid line is computed using actual non-ideal experimental parameters, which are measured independently.  The agreement between the calculation and the experimental data is excellent.

%%%%%%%%%%%%%%%%%%%%%%%%%%%%%%%%%%

A similar setup, but with significantly larger spatial and temporal separations, uses a 144~km free space link between the interferometer and the polarization projection setup (shown in Fig.\ \ref{canarysetup}). The two labs are located on two of the Canary Islands, La Palma and Tenerife~\cite{Ursin2007,Fedrizzi2009,Scheidl2010}. Two different space-time arrangements are realized, one of which achieves space-like separation of all relevant events. Within this scenario, different times for the choice events are chosen.  One arrangement is such that the speed of a hypothetical superluminal signal from the choice event \textbf{C}$_{e}$ to the events related to the interferometer \textbf{I}$_{s}$ would have to be about 96 times the speed of light, ruling out an explantation by prorogation influence ~\cite{Salart2008}.  The other arrangement is such that the choice event \textbf{C}$_{e}$ happens approximately 450~$\mu$s after the events \textbf{I}$_{s}$ in the reference frame of the source, which puts a record to the amount of delay by more than 5 orders of magnitude comparing to the previously reported quantum eraser experiment~\cite{Kim2000}. Even though the signal-to-noise ratio is reduced due to the attenuation of the free space link (33~dB), results similar to the Vienna experiment are obtained after subtraction of the background. See supplementary information for details.

Furthermore, for all the data obtained in the Vienna and Canary experiments, in order to achieve complete independence between the data registration of the system photon and the environment photon, we use two time-tagging units and individually record the time stamps of their detection events. These data are compared and sorted to reconstruct the coincidence counts, long after the experiment is finished~\cite{Weihs1998}.

Our work demonstrates and confirms that whether the correlations between two entangled photons reveal welcher-weg information or an interference pattern of one (system) photon, depends on the choice of measurement on the other (environment) photon, even when all the events on the two sides that can be space-like separated, are space-like separated. The fact that it is possible to decide whether a wave or particle feature manifests itself long after---and even space-like separated from---the measurement teaches us that we should not have any naive realistic picture for interpreting quantum phenomena. Any explanation of what goes on in a specific individual observation of one photon has to take into account the whole experimental apparatus of the complete quantum state consisting of both photons, and it can only make sense after all information concerning complementary variables has been recorded. Our results demonstrate that the view point that the system photon behaves either definitely as a wave or definitely as a particle would require faster-than-light communication. Since this would be in strong tension with the special theory of relativity, we believe that such a view point should be given up entirely.

\section*{Acknowledgement}
We are grateful to \v{C}. Brukner, M. \.{Z}ukowski, M. Aspelmeyer and N. Langford for helpful discussions as well as to T. Bergmann and G. Mondl for assistance with the electronics. We also wish to thank F. Sanchez (Director IAC) and A. Alonso (IAC), T. Augusteijn, C. Perez and the staff of the Nordic Optical Telescope (NOT), J. Kuusela, Z. Sodnik and J. Perdigues of the Optical Ground Station (OGS) as well as J. Carlos and the staff of the Residence of the Observatorio del Roque de Los Muchachos for their support at the trial sites. We acknowledge support from the European Commission, Project QAP (No.015848), Q-ESSENCE (No. 248095), ERC Senior Grant (QIT4QAD), the Marie-Curie research training network EMALI, and SFB-FOQUS, Templeton Fellowship at the Austrian Academy of Science's Institute for Quantum Optics and Quantum Information Vienna, and the Doctoral Program CoQuS of the Austrian Science Foundation (FWF).

\section{Supplementary information}
\subsection{Details of the Vienna experiment}
\textit{Entangled photon source.} In the Vienna experiment, a picosecond-pulsed Nd:Vanadate laser emitting light at 355~nm wavelength with a repetition rate of 76 MHz and an average power of 100 mW pumps a $\beta $-barium borate ($\beta $-BBO) crystal in a type-II scheme of spontaneous parametric down conversion (SPDC).  The coincidence rate is about 5~kHz and the single count rate is about 50~kHz.  Without subtraction of the background, we observe polarization correlations in the H/V basis with a visibility of 98.0\% and in the R/L basis with a visibility of 96.9\%, where H/V and R/L indicate horizontal/vertical and right/left circular polarization.

\textit{Polarization projection setup.} We use an electro-optical modulator (EOM) based on a Rubidium Titanyl Phosphate (RTP) crystal to switch between the H/V and R/L bases. The optical axis of the RTP is aligned to 45$^\circ$ such that the EOM leaves the input state unchanged for zero voltage and acts as a QWP at 45$^\circ$ when positive quarter voltage is applied. An FPGA logic samples the random bit sequence from the Quantum Random Number Generator (QRNG) and controls the EOM driver. For a random bit value of `1', +QV (770~V) is applied, for `0' the EOM is switched back to 0~V. The rise time of the EOM is 4.5~ns and the switching on time of each +QV cycle is about 20 ns. The toggle frequency is 2 MHz. The resulting duty cycle is hence approximately 20 ns $\times$ 2 MHz = 4$\%$.

\textit{Quantum random number generator (QRNG).} The total amount of the delay occurring in the electronics and optics of our QRNG is measured to be 75~ns.  Allowing for another 33~ns (3 times the autocorrelation time of QRNG), to be sure that the autocorrelation of the QRNG output signal is sufficiently low, the total duration of choice-related events is about 108~ns.  This is the reason why the choice event \textbf{C}$_{e}$ is a series of events instead of a single well-defined event in Fig.\ 2\textbf{B} in the main text. The temporal center of these events are chosen to be simultaneous with the emission event, \textbf{E}$_{se}$. In total, we thus exclude any influences between the measurements of the two photons as well as between the choice event and the system photon, at a speed equal to or less than speed of light.  It is important to address that, since \textbf{C}$_{e}$ is space-like separated from \textbf{E}$_{se}$, we also exclude any causal influence between the QRNG and the photon source~\cite{Bell2004sup,Scheidl2010sup}.

\textit{Space-time arrangements for the Vienna experiment.} The space-like separation between \textbf{I}$_{s}$ and \textbf{P}$_{e}$ is achieved in two steps. First, after the generation of the photon pairs from the source in Lab~1, we send the environment photon via a 55~m (275~ns) single-mode fibre to the polarization measurement setup in Lab~2 located 50~m away from Lab~1 (straight-line distance), as shown in Fig.\ 2\textbf{A} in the main text. In Lab~2 two polarization measurements (i) and (ii), described in the main text, are implemented by a fast EOM followed by a polarizing beam splitter. In measurement (i) the EOM is switched off, and in measurement (ii) the EOM is switched on by applying the quarter-wave voltage. Second, the system photon is delayed with a 28~m (140~ns) single-mode fibre in Lab~1 and sent into a 2~m (10~ns) fibre based interferometer. With this arrangement, \textbf{P}$_{e}$ is not only delayed with respect to \textbf{I}$_{s}$ in the laboratory frame of reference, but also sits outside of the past and the future light cones of \textbf{I}$_{s}$, as shown in Fig.\ 2\textbf{B} in the main text.

To fulfil the space-like separation between \textbf{C}$_{e}$ and \textbf{I}$_{s}$ as well as \textbf{C}$_{e}$ and \textbf{E}$_{se}$, the measurement basis is randomly determined by the QRNG located in Lab~3 with 27~m straight-line distance to Lab~2 and 77~m to Lab~1.  The random bits are transmitted via a 28~m (140~ns) coaxial cable to Lab 2. After the controller of EOM, there are about 7~m (35~ns) coaxial cables next to the EOM.

\begin{figure}[ht]
  \begin{center}
  \centerline{\includegraphics[width=1\textwidth]{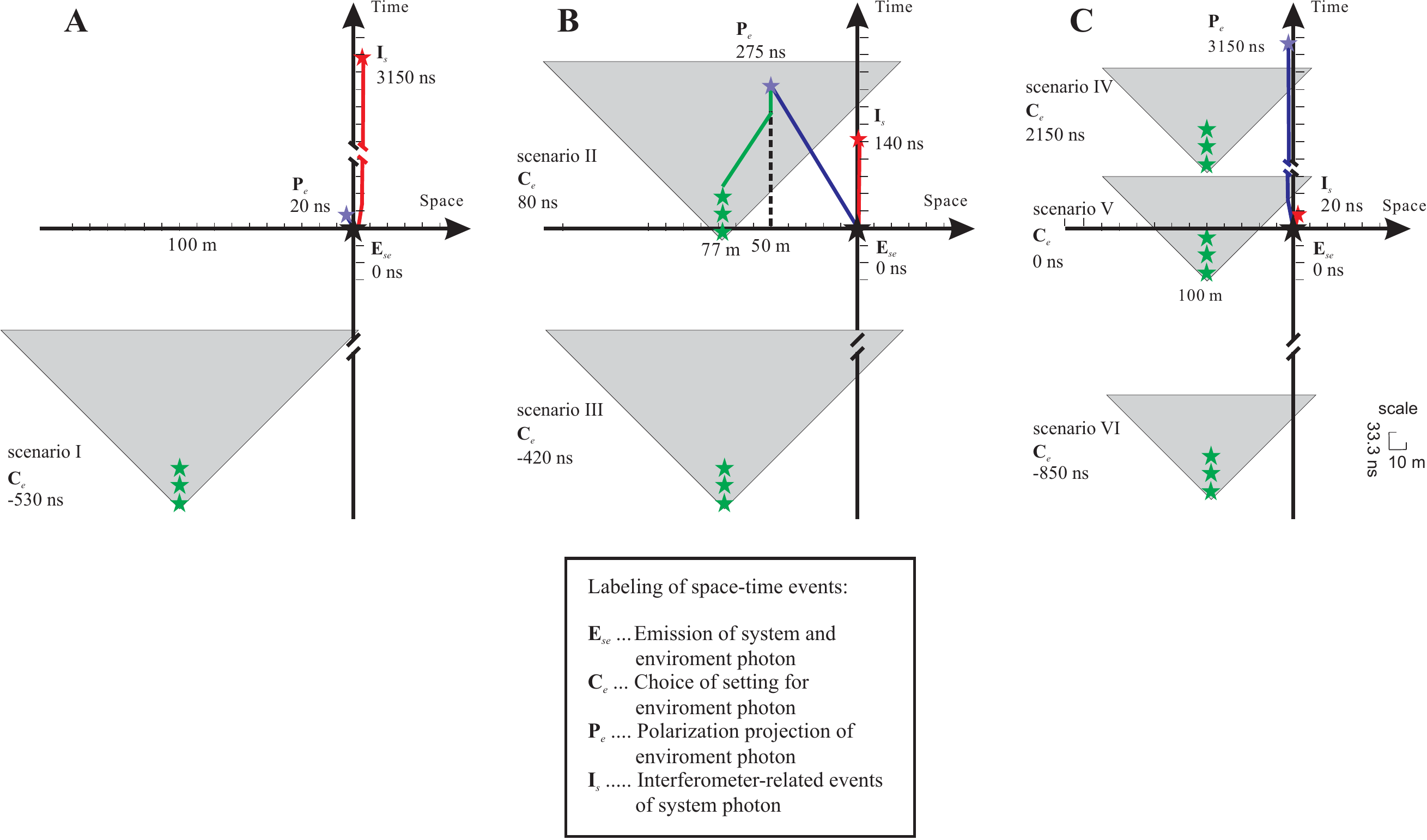}}
\caption{Six space-time scenarios arranged in the Vienna experiment are illustrated with the space-time diagrams in \textbf{A}, \textbf{B} and \textbf{C}. The green stars stand for the choice-related events \textbf{C}$_{e}$ and the grey light cone for the future light cone of the choice.  The polarization projection of the environment photon \textbf{P}$_{e}$ according to the choice, all
events related to the system photon inside the interferometer \textbf{I}$_{s}$, and the emission event of the photon pairs \textbf{E}$_{se}$ are blue, red and black stars, respectively.  Each unit of the space and time axes stands for 10~m and 33.3~ns respectively. Scenarios I, III, IV to VI are not scaled because of the large amount of delay on either the system or the environment photon side.} \label{STVienna}
  \end{center}
  \end{figure}

In the optical fibre link experiment performed in the campus of the University of Vienna, we demonstrate six different space-time scenarios of the four events mentioned above.  The space-time diagrams of the six scenarios are shown in Fig.\ \ref{STVienna} and the relations between each event together with the experimental results of each scenario are summarized in Tab.\ \ref{tabVienna}.  All six scenarios give the same results up to the statistical error.

Experimentally, we use optical fibres and coaxial cables to delay and distribute the optical and electrical signals in order to fulfil the requirement of each scenario.  For instance, in order to achieve scenario V, we arrange the location of each apparatus in the following way.  The environment photon is delayed locally (close to the source and interferometer) with a 630~m single-mode fibre, and there is no delay on the system photon side.  The straight-line distance between the QRNG and the interferometer is about 100~m.  The random bits are transmitted with a 630~m coaxial cable to the EOM. The other scenarios are realized in similar ways by changing the location of the polarization projection setup of the environment photon and the length of the fibre and coaxial cable.

\begin{table}[h]
\begin{tabular}[c]{|c|p{0.45in}|p{0.45in}|p{0.45in}|p{0.49in}|p{0.45in}|p{0.45in}|p{0.45in}|p{0.45in}|p{0.45in}|m{0.65in}|m{0.65in}|} \hline
\textbf{Scenarios} & \multicolumn{9}{c|}{\textbf{Relations between events}} & \multicolumn{2}{c|}{\textbf{Results}} \\ \hline
 & \textbf{P}$_{e}$ before \textbf{I}$_{s}$ & \textbf{P}$_{e}$ s.l.\ sep.\ \textbf{I}$_{s}$ & \textbf{P}$_{e}$ after \textbf{I}$_{s}$ & \textbf{I}$_{s}$ before \textbf{C}$_{e}$ & \textbf{I}$_{s}$ s.l.\ sep.\ \textbf{C}$_{e}$ & \textbf{I}$_{s}$ after \textbf{C}$_{e}$ & \textbf{E}$_{se}$ before \textbf{C}$_{e}$ & \textbf{E}$_{se}$ s.l.\ sep.\ \textbf{C}$_{e}$ & \textbf{E}$_{se}$ after \textbf{C}$_{e}$ & \multicolumn{1}{c|}{$\mathcal{I}_\textrm{(i)}$}  & \multicolumn{1}{c|}{$\mathcal{V}_\textrm{(ii)}$} \\ \hline
I & \multicolumn{1}{c|}{{X}} &   &   &   &  & \multicolumn{1}{c|}{X} & \textbf{} & \textbf{} & \multicolumn{1}{c|}{X} & \multicolumn{1}{c|}{0.956(8)} & \multicolumn{1}{c|}{0.950(20)} \\ \hline
II & \textbf{} & \multicolumn{1}{c|}{{X}} & \textbf{} & \textbf{} & \multicolumn{1}{c|}{{X}} & \textbf{} & \textbf{} & \multicolumn{1}{c|}{{X}} & \textbf{} & \multicolumn{1}{c|}{0.955(7)} & \multicolumn{1}{c|}{0.951(18)} \\ \hline
III & \textbf{} & \multicolumn{1}{c|}{{X}} & \textbf{} & \textbf{} & \textbf{} & \multicolumn{1}{c|}{{X}} & \textbf{} & \textbf{} & \multicolumn{1}{c|}{{X}} & \multicolumn{1}{c|}{0.953(9)} & \multicolumn{1}{c|}{0.952(18)} \\ \hline
IV & \textbf{} & \textbf{} & \multicolumn{1}{c|}{{X}} & \multicolumn{1}{c|}{{X}} & \textbf{} & \textbf{} & \multicolumn{1}{c|}{{X}} & \textbf{} & \textbf{} & \multicolumn{1}{c|}{0.957(7)} & \multicolumn{1}{c|}{0.946(21)} \\ \hline
V & \textbf{} & \textbf{} & \multicolumn{1}{c|}{{X}} & \textbf{} & \multicolumn{1}{c|}{{X}} & \textbf{} & \textbf{} & \multicolumn{1}{c|}{{X}} & \textbf{} & \multicolumn{1}{c|}{0.957(7)} & \multicolumn{1}{c|}{0.943(21)} \\ \hline
VI & \textbf{} & \textbf{} & \multicolumn{1}{c|}{{X}} & \textbf{} & \textbf{} & \multicolumn{1}{c|}{{X}} & \textbf{} & \textbf{} & \multicolumn{1}{c|}{{X}} & \multicolumn{1}{c|}{0.954(8)}& \multicolumn{1}{c|}{0.950(19)} \\ \hline
\end{tabular}
\caption{Summary of the space-time relations and experimental results for the six scenarios performed in the Vienna experiment.  There are three different possible relations.  For the events \textbf{P}$_{e}$ and \textbf{I}$_{s}$, e.g., there are the relations `\textbf{P}$_{e}$ before \textbf{I}$_{s}$', `\textbf{P}$_{e}$ s.l.\ sep.\ \textbf{I}$_{s}$' and `\textbf{P}$_{e}$ after \textbf{I}$_{s}$', which mean that (in all reference frames including the lab frame) event \textbf{P}$_{e}$ happens `time-like before' (i.e. \textbf{P}$_{e}$ is in the past light cone of \textbf{I}$_{s}$), `in a space-like separated region with respect to' and `time-like after' (i.e. \textbf{P}$_{e}$ is in future light cone of \textbf{I}$_{s}$) event \textbf{I}$_{s}$, respectively.  There is an `X' when the relation is fulfilled in the corresponding scenario. When measurement (i) is performed, almost full welcher-weg information of the system photon is acquired.  When measurement (ii) is performed, high visibility interference fringes show up because almost the complete welcher-weg information is erased.}\label{tabVienna}
\end{table}

\textit{Independent data registration.} To achieve complete independence between the events related to the environment photon and those related to the system photon, one should avoid the registration of coincidences between the photon pairs with the same electronic unit. More preferably, the individual events should be registered on both sides independently and compared only after the measurements are finished~\cite{Weihs1998sup}.  In our experiment, independent data registration is performed with two time-tagging units.  In the Vienna experiment, the time bases on both sides are established by using a function generator with 10 MHz signal outputs as local clocks and another function generator with 1 Hz signal outputs to provide the time reference for synchronizing both sides' time tags.  On the system photon side, the pulses of Det~1 and Det~2 are directly fed into one time-tagging unit.  On the environment photon side, the detection signals of Det~3 and Det~4 with the bit values from the QRNG are first fed into a field-programmable gate array (FPGA) logic, responsible for distinguishing whether the EOM is on or off.  Finally, all detections on the environment photon side are recorded by the other time-tagging unit referenced to the clock together with the corresponding EOM status.  Each side features a personal computer which stores the tables of time tags accumulated and on the system photon side the corresponding position of the phase scanner is also recorded.  Long after the measurements are finished, coincidence counts are identified by calculating the time differences between Alice's and Bob's time tags and comparing these within a time window of 1~ns.

\textit{The experimental complementarity inequality.} Experimentally, there are several practical limitations for obtaining perfect complementarity.  The measurement of the welcher-weg information parameter, $\mathcal{I}$, is limited by the correlation in the H/V basis (98.0\%) and the imperfection of the PBS (extinction ratio 180:1).  These imperfections are taken into account by a correction factor $\eta _{\mathcal{I}} $ of about 0.97.  On the other hand, $\mathcal{V}$ is limited by the correlation in the R/L basis (96.9\%), the imperfection of the PBS and the imperfection of polarization rotation of the EOM (extinction ratio 250:1).  These imperfections are taken into account by a correction factor $\eta _{\mathcal{V}} $ of about 0.95.  So the inequality becomes: $\frac{\mathcal{I}^{2} }{\eta _{\mathcal{I}}^{2} } +\frac{\mathcal{V}^{2} }{\eta _{\mathcal{V}} ^{2} } \le 1$, which can be rewritten as:
\begin{equation} \label{GrindEQ__3_}
\mathcal{V}\le \eta _{\mathcal{V}} \sqrt{1-\frac{\mathcal{I}^{2} }{\eta _{\mathcal{I}} ^{2} } }
\end{equation}
When all the other imperfections are excluded, the upper limit is reached.  As we expected, the experimental data agree well with the function of $\mathcal{V}=0.95\sqrt{1-\left(\mathcal{I}/0.97\right)^{2}}$.

\subsection{Details of the Canaries experiment}

\textit{Setup of the Canaries experiment.} We perform the free-space link experiment between La Palma and Tenerife, two Canary Islands off the West African coast with a straight-line distance of about 144~km.  This corresponds to a photon flight time of about 479~$\mu$s. The optical free-space link is formed by a transmitter telescope mounted on a motorized platform and a receiver telescope, the European Space Agency's optical ground station (OGS) with a 1~m mirror (effective focal length  \textit{f }= 38~m) located on Tenerife. The transmitter consists of a single-mode fibre coupler and an \textit{f}/4 best form lens (\textit{f }= 280~mm). The employed closed-loop tracking system is described in refs.\ \cite{Ursin2007sup, Fedrizzi2009sup, Scheidl2010sup}. Using a weak auxiliary laser diode at 810~nm, the attenuation of the optical link, starting from the 10~m single-mode fibre to the transmitter telescope to the APD in Tenerife with an active area of about 500~$\mu$m in diameter at the OGS is measured to be about 35~dB. The photon pair attenuation through the whole setup is therefore 3~dB + 35~dB = 38~dB, where the 3~dB attenuation is due to the 1 km fibre delay on the system photon side. In Fig.\ 4 of the main text, we show an overview of the experimental scheme.

In this experiment, entangled photon pairs are generated via SPDC in a 10 mm ppKTP crystal which is fabricated for type-II phase-matching condition and is placed inside a polarization Sagnac interferometer. Implementing a 405~nm laser diode with a maximum output power of 50 mW, we are able to generate entangled pairs with a production rate of 3.4$\times$10$^\textrm{7}$ Hz. This number is inferred from locally detected 250000 photon pairs/s at a pump power of 5 mW and a coupling efficiency of 27\%. Furthermore, operation at 5 mW pump power yields a locally measured visibility of the generated entangled state in the H/V (+/-) basis of about 99\% (98\%) (accidental coincidence counts subtracted), and we assume that the state visibility does not change considerably at 50 mW pump power. Note that ${\left| \textrm{+} \right\rangle}$ and ${\left| \textrm{--} \right\rangle}$ are the 45$^\circ$ and --45$^\circ$ polarization states (with ${\left| \textrm{+} \right\rangle} =({\left| \textrm{H} \right\rangle} +{\left| \textrm{V} \right\rangle} )/\sqrt{2} $ and ${\left| \textrm{--} \right\rangle} =({\left| \textrm{H} \right\rangle} -{\left| \textrm{V} \right\rangle} )/\sqrt{2} $).

In the Canaries experiment, we align the optical axes of the RTP crystals to 22.5$^\circ$. Additionally, we place a QWP with its optical axis oriented parallel to the axis of the RTP crystals in front of the EOM. Applying positive quarter-wave voltage (+QV) makes the EOM act as an additional QWP, such that the overall effect is like a HWP at 22.5$^\circ$ which rotates the polarization by 45$^\circ$. On the contrary, applying negative quarter-wave voltage (--QV) makes the EOM compensate the action of the QWP, such that the overall polarization rotation is 0$^\circ$.  A random bit `0' (`1') requires that a polarization rotation of  0$^\circ$ (45$^\circ$) and --QV (+QV) is applied to the EOM.  A certain setting is not changed until the occurrence of an opposite trigger signal. However, since our QRNG is balanced within the statistical uncertainties, +QV and --QV are applied close to equally often. A toggle frequency of about 1 MHz is used. The rise time of the EOM is measured to be $<$15 ns, thus to be sure that the switching process has been finished, we discard all photons which are detected less than 35~ns after a trigger signal. This kind of operation results in a switching duty cycle of approximately 96.5\%.

For the Canaries experiment, similar electronics are employed except the synchronization of the time bases of two time-tagging units. The time-tagging units are disciplined to the global positioning system (GPS) time standard.

\begin{figure}[hp!]
\centerline{\includegraphics[width=0.675\textwidth]{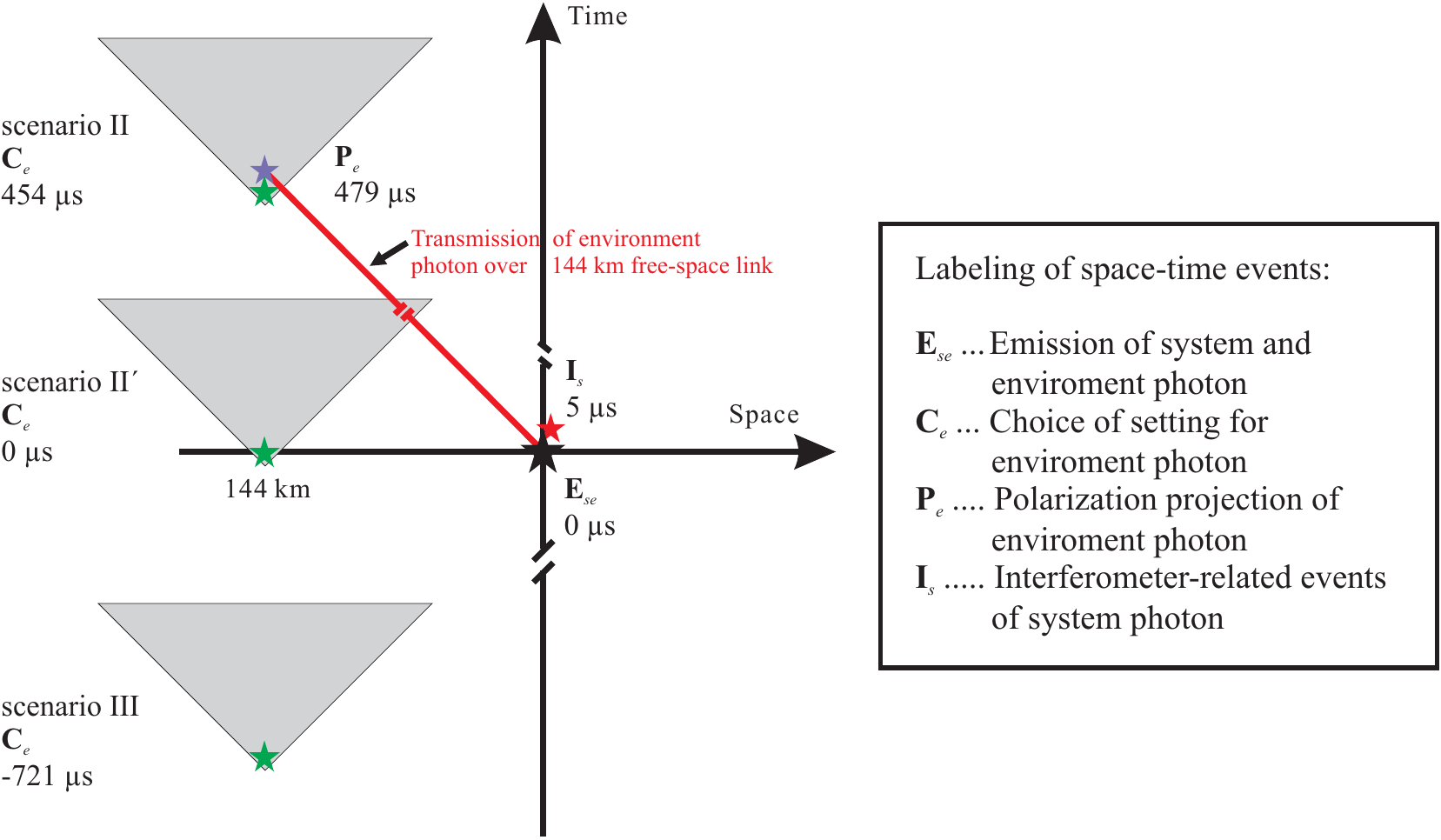}} \caption{The space-time diagrams of three experiments performed with the free-space link, which correspond to two different space-time scenarios. The labels of the events and light cones are the same as in Fig.\ \ref{STVienna} in the Supplementary Information for details. Scenarios II, II' and III are the arrangements where the event \textbf{C}$_{e}$ is later, at the same time and earlier than the event \textbf{E}$_{se}$ respectively in the lab reference frame.} \label{STCanary}
\end{figure}

\textit{Space-time arrangements for the Canaries experiment.} In the Canaries experiment, we demonstrate three different space-time arrangements of the four relevant events: \textbf{P}$_{e}$, \textbf{C}$_{e}$, \textbf{I}$_{s}$ and \textbf{E}$_{se}$.  The flight time of the environment photon from the lab in La Palma to the lab in Tenerife is about 479~$\mu$s. In the lab on La Palma, using an optical fibre, we introduce a delay to the system photon before its entry into the Mach-Zehnder interferometer.  The length of the delay fibre is 1 km, which corresponds to a photon flight time of about 5~$\mu$s.  This experimental arrangement space-like separates the polarization projection of the environment photon \textbf{P}$_{e}$ from the entry of system photon into the interferometer \textbf{I}$_{s}$ in a relativistic sense.  We keep this arrangement to be the same for all three scenarios. The relative time delays between the choice \textbf{C}$_{e}$ and \textbf{E}$_{se}$ are adjusted to be --721~$\mu$s, 0~$\mu$s and 454~$\mu$s for Scenario III, II' and II, respectively. The space-time diagrams of all three scenarios are shown in Fig.\ \ref{STCanary} and the relations between all events are summarized in Tab.\ \ref{tableCanary}.  These three scenarios gave similar results up to statistical errors.

\begin{table}
\begin{tabular}[c]{|c|p{0.45in}|p{0.45in}|p{0.45in}|p{0.45in}|p{0.45in}|p{0.45in}|p{0.45in}|}\hline
\textbf{Scenarios} & \multicolumn{5}{c|} {\textbf{Relations between events} } & \multicolumn{2}{c|}{\textbf{Results} } \\ \hline
 & \textbf{P}$_{e}$ s.l.\ sep.\ \textbf{I}$_{s}$  & \textbf{I}$_{s}$ s.l.\ sep.\ \textbf{C}$_{e}$  & \textbf{I}$_{s}$ after A  & \textbf{E}$_{se}$ s.l.\ sep.\ \textbf{C}$_{e}$  & \textbf{E}$_{se}$ after \textbf{C}$_{e}$  & \multicolumn{1}{c|}{$\mathcal{I}_\textrm{(i)}$}  & \multicolumn{1}{c|}{$\mathcal{V}_\textrm{(ii)}$}  \\ \hline
II  & \multicolumn{1}{c|}{{X}}  & \multicolumn{1}{c|}{{X}}  &  & \multicolumn{1}{c|}{{X}}  &  & \multicolumn{1}{c|}{0.932(2)}  & \multicolumn{1}{c|}{0.756(2)}  \\ \hline
II'  & \multicolumn{1}{c|}{{X}}  & \multicolumn{1}{c|}{{X}}  &  & \multicolumn{1}{c|}{{X}}  &  & \multicolumn{1}{c|}{0.930(2)}  & \multicolumn{1}{c|}{0.776(2)}  \\ \hline
III  & \multicolumn{1}{c|}{{X}}  &  & \multicolumn{1}{c|}{{X}}  &  & \multicolumn{1}{c|}{{X}}  & \multicolumn{1}{c|}{0.928(1)}  & \multicolumn{1}{c|}{0.751(1)}  \\ \hline
\end{tabular}
\caption{Summary of the space-time relations between events for three scenarios of the Canaries experiment. The conventions of the relationship between different events are the same as that in Tab.\ \ref{tabVienna}. When measurement (i) is performed, almost full welcher-weg
information of the system photon is acquired.  When measurement (ii) is
performed, interference with high visibility shows up because the welcher-weg information is erased.} \label{tableCanary}
\end{table}

\end{document}